# A Simple Experimental System for Predicting the Temperature Rise of the Global Warming Induced by the Greenhouse Effect

Yoshiyuki Kawamura

Department of Intelligent Mechanical Engineering, Fukuoka Institute of Technology, 3-30-1 Wajirohigashi, Higashiku, Fukuoka, 811-0295, Japan (kawamura@fit.ac.jp)

Abstract

In this report, a simple experimental system is shown, by which the temperature rise of global warming due to greenhouse gases can be demonstrated quantitatively. The system configuration is similar to that of the earth-atmosphere-space system based on a simple gray atmosphere model. The space (heat sink), the ground surface and the radiation power from the sun were replaced by a liquid nitrogen, and a black coated aluminum plated with an electric heater and an electric power of the heater, respectively. We believe that it can be a unique experimental system, that is useful for the study of the greenhouse effect in a laboratory, as well as for the educational demonstration tool of this phenomenon.

# 1. Introduction

Global warming is one of the most important problems facing humanity, and the temperature rise due to the greenhouse effect is the key to understanding this phenomenon [1, 2]. Global warming has been studied using geoscientific [3, 4, 5] and meteorological approaches [6, 7]. Among various types of greenhouse gases, Methane is important to study, because its concentration in the atmosphere has approximately doubled over the last century and continues to increase exponentially [8].

To predict how much the global temperature will rise because of the increase in greenhouse gas concentration, such as carbon dioxide gas and methane gas, a simple laboratory-scale experimental system simulating the earth-atmosphere-space system [9] based on a simple gray atmosphere model [6] was developed. In this experiment, space (the heat sink), the ground surface, and the radiation power from the sun were simulated by liquid nitrogen, a radiation plate (aluminum-plated electric heater painted black), and the electrical power of the heater, respectively. The temperature of the radiation plate was measured as a function of the methane gas concentrations.

The experimental results obtained by this system suggest that doubling the current concentration of atmospheric methane will increase the average global temperature by 1.3 K.

# 2. Experimental system

Figure 1 shows the experimental setup for the measurement of the temperature rise as a function of methane gas concentration. A gas cell 0.55 m in depth and 0.3 m × 0.3 m in cross-section was filled with methane. The total pressure of the mixed gas was 1 atm. The concentration of methane in the cell can be changed by adding nitrogen gas. The inner wall of the gas cell was covered with aluminum to reflect infrared radiation into the cell.

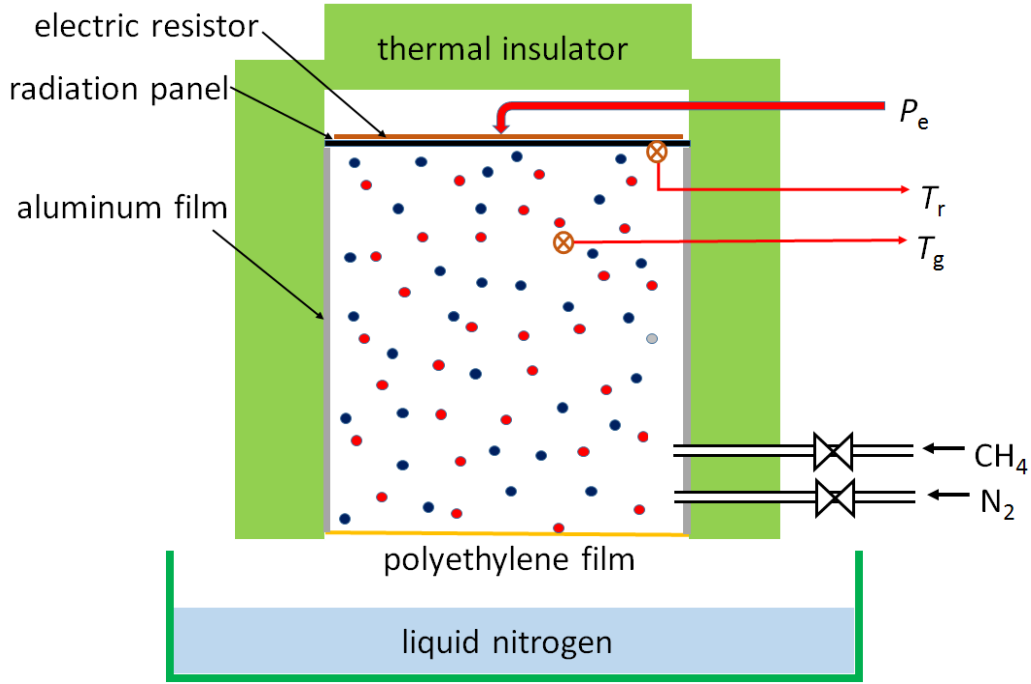

Fig. 1 Experimental setup for simulating the temperature rise caused by the greenhouse effect.

A thin aluminum plate (0.3m × 0.3m × 100 μm) was placed on the top of the gas cell. The surface of the plate in contact with the gas cell was painted black. The plate was heated by a thin wire heater with constant electrical power $P_e$, which represents the power radiated by the sun onto the ground. The electrical power was calculated to be

$$P_e = 1370 \text{ W/m}^2 \times \frac{168 \text{W/m}^2}{342 \text{W/m}^2} \times \frac{\pi R^2}{4\pi R^2} \times 0.09 \text{ m}^2 = 15\text{W} \qquad (1),$$

where the first term is a solar constant, the second term is the ratio of the radiation power density on the ground to that at the top of the atmosphere [10], the third term is the ratio between the cross-section and the surface area of the earth, the fourth term is the cross-section of the gas cell, and $R$ is the radius of the earth.

The bottom of the gas cell was sealed using a 15-μm polyethylene thin film. This material was found to have a relatively high average transmissivity of 85% between 400 cm⁻¹ and 2000 cm⁻¹, and it had sharp absorption lines near 700 cm⁻¹ and 1500 cm⁻¹ (Fig. 2). The sinusoidal waveform observed between 400 cm⁻¹ and 2000 cm⁻¹ was produced by the interference between the reflections from opposite surfaces of the film.

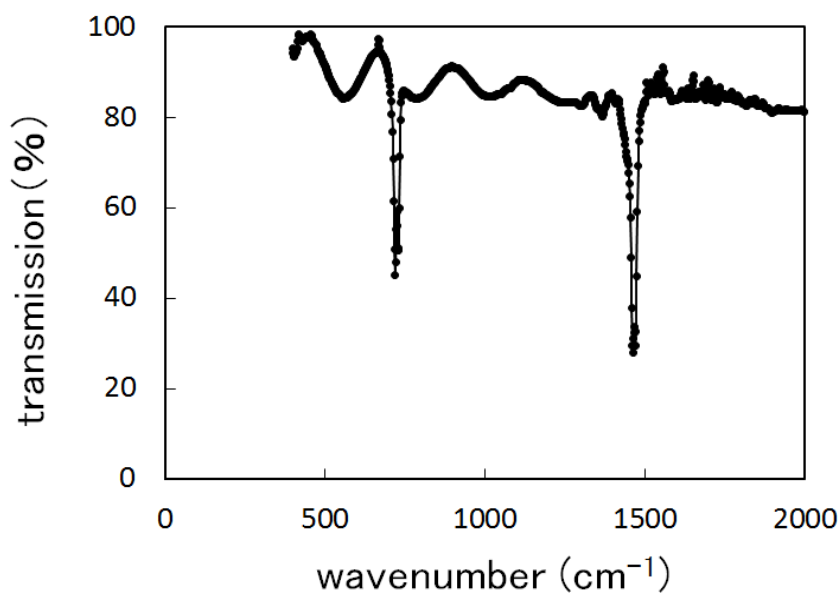

Fig. 2 Transmission of a polyethylene thin film with a thickness of 15 $\mu$m.

    Liquid nitrogen was placed on the bottom of the gas cell as a heat sink to absorb the radiation from the radiation panel. Although the temperature of the liquid nitrogen (77 K) is higher than the cosmic temperature (3 K), it was low enough for this experiment because the radiant power is proportional to the fourth power of the radiation temperature.

    The side walls of the gas cell and the upper side of the radiation plate were covered by thermal insulators (extruded polystyrene foam) with a thickness of 0.1m and a thermal conductivity coefficient of 0.036W/m²K to minimize the heat loss to the surroundings. The temperature of the radiation plate ($T_b$) and the gas ($T_g$) in the cell were measured using thermocouple thermometers. Due to the limited tensile strength of the polyethylene thin film, the total pressure in the gas cell was chosen to be atmospheric pressure.

The real atmosphere is equivalent to an air column of length 8015 m at a pressure of 1 atm. Methane, which is currently present in the atmosphere at a concentration of 1.87 ppm, is therefore equivalent to a column length of 0.015 m. In this experiment, the length of the gas cell was 0.55 m. The current concentration of methane in the atmosphere (1.87 ppm) corresponds to a methane concentration of 2.7% (0.015 m/0.55 m) in the gas cell, and the expected future doubling of the methane concentration corresponds to 5.4% methane in the gas cell.

## 3. Experimental results and discussion

Figure 3 shows the temperature of the radiation plate as a function of the methane gas concentration. It is seen that the radiation plate temperature increases, as the methane gas concentration increases. The green and red arrows correspond to the present methane concentration (1.87 ppm in the atmosphere, 2.7% in the cell) and the doubled methane concentrations (3.74 ppm in the atmosphere, 5.4% in the cell), respectively. Doubling the methane concentration results in a temperature increase of about 1.3 K of the radiation plate temperature.

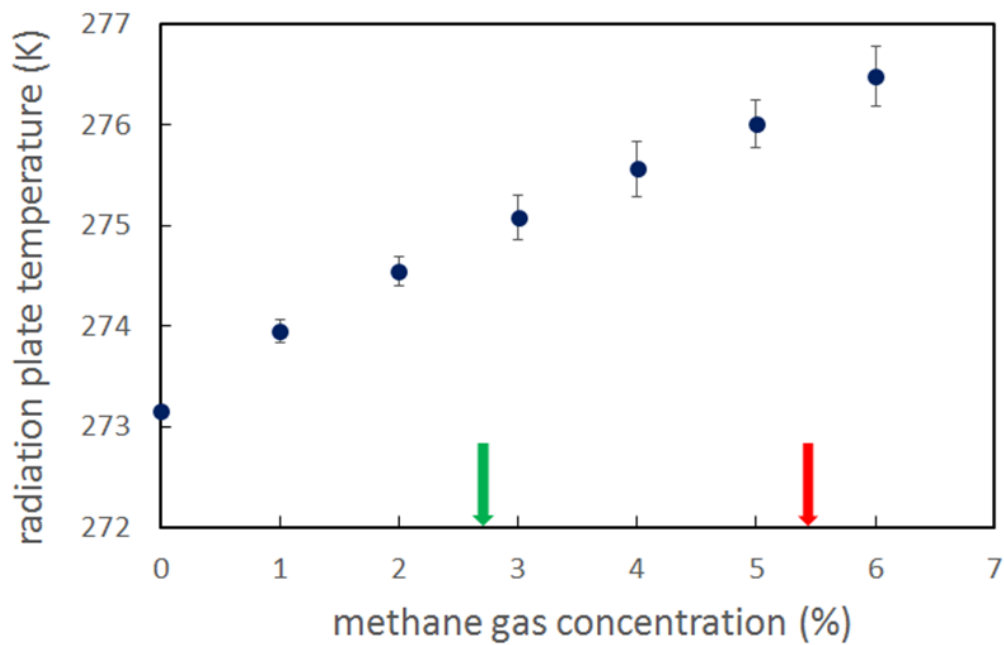

Fig. 3 Radiation panel temperature as a function of methane gas concentration. The green and red arrows represent the present and the doubled methane concentrations.

It is well known that the temperature increase is also influenced by water vapor ($H_2O$). The absorption band spectrum of the methane has a central wavenumber of about 1300 cm$^{-1}$ and a spectral width of about 150 cm$^{-1}$. We used a transmission spectral curve from a standard textbook and a calculation tool for the infrared spectrum [6, 11] to determine that the average transmissivity of water vapor in this spectral range is 53%.

The temperature rise is proportional to the radiative forcing. Considering that the radiative forcing is attenuated by the water vapor, the temperature rise is corrected to be about 0.7 degrees.

## 4. Conclusions

We developed a simple laboratory-scale experimental system to predict the global temperature rise caused by greenhouse gases in the atmosphere. The system is analogous to the actual earth-atmosphere-space system and is based on a simple gray-atmosphere model. It was shown that doubling the methane concentration resulted in a temperature rise of about 1.3 K. We believe that it can be a unique experimental system, which is used for the study on the greenhouse effect in a laboratory, as well as for the educational demonstration tool of this phenomena.